\documentclass[doublecol]{epl2}
\def\nabp{\nabla_{\perp}}
\newcommand{\sss}{\scriptscriptstyle}
\def\vti{v_{{\sss T}i}}
\def\vte{v_{{\sss T}e}}
\newcommand {\be}{\begin{equation}} % start equation
\newcommand{\ee}{\end{equation}}    % end equation

\title{A new paradigm for solar coronal heating}
\shorttitle{Title} %Insert here a short version of the title if it exceeds 70 characters

\author{J. Vranjes
\and S. Poedts
} \shortauthor{J. Vranjes and S. Poedts}

\institute{
  K. U. Leuven, Center for Plasma Astrophysics,
Celestijnenlaan 200B, 3001 Leuven,
 Belgium, and\\ Leuven Mathematical Modeling and Computational Science Center
 (LMCC)\\
}

\abstract{The solar coronal heating problem refers to the question
why the temperature of the Sun's corona is more than two orders of
magnitude higher than that of its surface. Almost 70 years after the
discovery, this puzzle is still one of the major challenges in
astrophysics. The current basic paradigm of coronal heating is
unable to explain all the observational features of the heating.
Here we argue that a new paradigm is required to solve the puzzle in
a self-consistent manner. The alternative approach is based on the
kinetic theory of drift waves. We  show, with qualitative and
quantitative arguments, that the drift waves  have the potential to
satisfy all coronal heating requirements.}

\begin{document}

\maketitle

%\section{Section title}

A self-consistent coronal heating model must fulfil a lot of
requirements imposed by observational facts. First of all, it should
be consistent with the measured energy losses in the solar corona
due to conduction and radiation, i.e.\ it should a)~not only provide
the right amount of energy but b)~do so at the right times scales,
e.g.\  $ \simeq 10^{-4}$ J/(m$^3$\,s) in active regions. Moreover,
it should c)~include the source of the required energy \cite{1}, and
d)~work everywhere in the corona, i.e.\ for all different magnetic
structures (with different heating requirements). Furthermore, it
should be able e)~to explain the observed temperature anisotropy
\cite{2,3} ($T_{\bot}
> T_{\|}$), f)~be more effective on ions than on electrons ($T_i >
T_e$), and g)~heat heavier ions more efficiently than lighter ions
\cite{4}. None of the proposed heating mechanisms so far even
claimed to fulfil all these model requirements.

The  current paradigm of coronal heating states that the required
energy source is provided by the plasma flows below the solar
surface and that this energy is transferred to the corona through
the motions of the `foot points' of the coronal magnetic field lines
which are `anchored' in this zone. Depending on the ratio $\tau$ of
the time scale of these `driving' motions to the dynamic (Alfv\'en)
time scale, the current models are classified as `wave heating'
($\tau<1$) or `magnetic reconnection (or nanoflare)' ($\tau>1$)
models \cite{5}. The main challenge, however, is to explain how the
energy is dissipated in the highly conductive corona with a
Lundquist number (the ratio of the dissipation time to the Alfv\'en
time scale) around $10^{13}$. All proposed mechanisms either have a
problem with the energy transport to the corona (the observed wave
fluxes are too small) or with the dissipation (too little and/or too
slow, or only sporadic) \cite{5}.

Most current models rely on the continuum or fluid approximation
(Magnetohydrodynamics, MHD). However, these models cannot really
explain coronal heating completely because i)~it is clear that the
actual heating takes place at length scales much smaller than those
on which the (macroscopic) MHD model is justified; and ii)~it is
obvious that the observed discrepancy between ion and electron
temperatures in the corona, as well as iii)~the observed large
temperature anisotropy in the inner corona ($T_{\bot} > T_{\|}$),
and iv)~the observed preferential heating of the heavier ions
\cite{2} are beyond the (single!) fluid model.

Here, we make a starting step towards the formulation of a new
paradigm, based on the   {\em kinetic theory of the drift waves}
driven by density gradients that are omnipresent in the solar
corona. It implies that the direct energy supply for the heating
comes from the corona itself (from the density gradients), though
still maintained and replenished by some mechanisms below the
surface. Those include a continuous restructuring of the magnetic
field, implying consequent similar changes of the plasma density
(due to the frozen-in conditions), and also the observed inflow of
the plasma along the magnetic loops \cite{6}. To some extent this
looks similar to the currently accepted scenarios mentioned above,
where the magnetic field plays an essential role and is  assumed as
given. However, the suggested new approach not only enables to
describe these drift waves (which are missing in the MHD picture),
but also their dissipation is easy to explain in the
\textit{self-consistent} kinetic model that works on the (very
small) length scales at which the actual dissipation takes place.
Actually, two mechanisms of energy exchange and heating will be
shown to take place simultaneously, one due to the Landau effect in
the direction parallel to the magnetic field, and another one, {\em
stochastic heating}, in the perpendicular direction. Moreover, this
stochastic drift wave heating mechanism seems to satisfy all seven
above-mentioned model requirements. This will be proved below using
only established basic theory, verified experimentally in laboratory
plasmas.

The {\em drift  mode} is the only mode that is not only able to
survive the drastically different  extremes in various parts of the
solar atmosphere but, in fact, even manages to benefit (i.e.\ to
grow) from each of them. The driving mechanism, however, is always
the same, viz.\  {\em the density gradients} perpendicular to the
ambient magnetic field vector. Numerous observations confirm the
omnipresence of such irregularities of the plasma density across
magnetic flux surfaces (see Fig.~1). Extremely fine density
filaments and threads have been observed in the solar atmosphere for
a long time now, even from ground-based observations \cite{7} like
those during the eclipse in 1991, showing a slow radial enlargement
of the structures. Contour maps \cite{8} reveal the existence of
numerous structures of various sizes. Filamentary structures with
length scales of the order of $1\;$km have been discussed \cite{9}.
Recent Hinode observations \cite{10}  confirm that the solar
atmosphere is a highly structured and very inhomogeneous system,
with radially spreading  density filaments of various size pervading
the whole domain. A very recent three-dimensional analysis of
coronal loops reveals short-scale density irregularities within each
loop separately \cite{11}. Such density irregularities are, as a
rule, associated with the magnetic field, thus creating a perfect
environment for drift waves. The characteristic dimensions of the
observed density irregularities are limited by the resolution of the
current instruments that is presently a fraction of an arcsec.
However, even extremely short, meter-size length scales can not be
excluded, especially in the corona \cite{12}. Hence, in dealing with
drift waves, we may operate with density inhomogeneity length scales
that have any value from one meter up to thousands of kilometers (in
the case of coronal plumes). Nevertheless, the role of the drift
wave in the problem of solar coronal heating has been overlooked  so
far in the literature, probably due to the fact that it simply does
not exist in the widely used MHD model.

\begin{figure}
\onefigure[height=8cm,bb=32 33 757
577,clip=,width=.85\columnwidth]{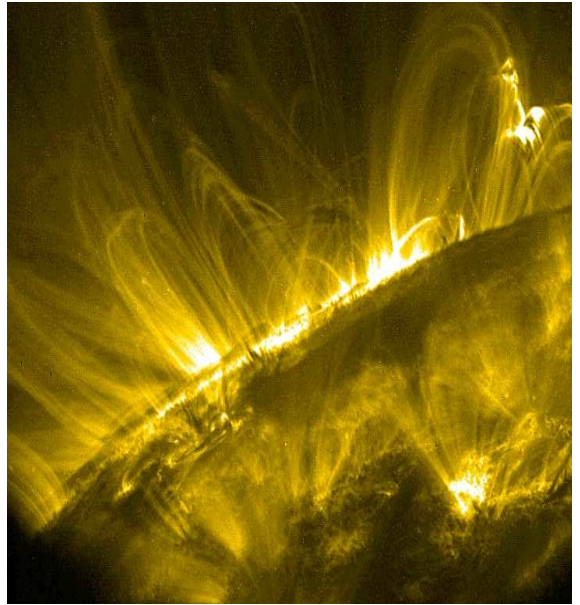} \caption{An image
of the Sun taken by the Transition Region and Coronal Explorer
(TRACE).} \label{fig.1}
\end{figure}

Regarding the wave heating analysis, in the case of the practically
collision-less corona, an efficient mechanism for the transfer of
energy from the wave to the plasma is needed. Within the framework
of the drift wave kinetic theory, this process develops as follows:
the interaction between the wave and the electrons is destabilizing
and the mode grows due to a Cherenkov-type interaction. At the same
time, however, its energy is absorbed by the ions due to Landau
damping. This may be seen from Ref.~\cite{13}, where the drift wave
properties within the limits $k_z \vti \ll\omega \ll k_z \vte$,
$\omega \ll \Omega_i$, and $|k_y/k_z| (T_e/T_i)^{1/2} \rho_i/L_n\gg
1$,  are described by the frequency
 \begin{equation}
\omega_r=-\frac{\omega_{*i} \Lambda_0(b_i)}{1- \Lambda_0(b_i) +
T_i/T_e + k_y^2 \lambda_{di}^2},\label{k1}
\end{equation}
and the growth rate
$$
\omega_i \simeq \left(\frac{\pi}{2}\right)^{1/2} \frac{\omega_r^2}
{\omega_{*i} \Lambda_0(b_i)}\left[\frac{T_i}{T_e} \frac{\omega_r -
\omega_{*e}}{|k_z|\vte} \exp[-\omega_r^2/(k_z^2 \vte^2)]\right.
$$
 \begin{equation}
  \left. +  \frac{\omega_r - \omega_{*i}}{|k_z|\vti}
\exp[-\omega_r^2/(k_z^2 \vti^2)]\right]. \label{k2} \end{equation}
Here, $ \Lambda_0(b_i)=I_0(b_i) \exp(-b_i)$,  $b_i=k_y^2 \rho_i^2$,
$\rho_i = \vti/\Omega_i$, $  \lambda_{di}=\vti/\omega_{pi}$,
$\omega_{*i}=- \omega_{*e} T_i/T_e$, $\omega_{*e}=k_y v_{*e}$, $\vec
v_{*e}=-(v_{{\sss T} e}^2/\Omega_e) \vec e_z \times \nabp n_0/n_0$,
and  $I_0$ is the modified Bessel function of the first kind and of
the order 0.

Eq. (\ref{k1}) reveals the energy source already in the real part of
the frequency $\omega_r\propto \nabla_\bot n_0$, while details of
its growth due to the same source are described by Eq. (\ref{k2}).
It is seen that, for a wave frequency below $\omega_{*e}$,  the
growth rate can be written  as
$\omega_i=|\gamma_{el}|-|\gamma_{ion}|$. The energy exchange between
the plasma particles and the wave is provided by {\em two
mechanisms} simultaneously, viz.\ one in the parallel  and one in
perpendicular direction.   The  term $|\gamma_{ion}|$ is responsible
for the Landau dissipation of the wave energy and, consequently, for
the parallel heating of the plasma. So, as long as the density
gradient is present, there is a continuous precipitation of energy
from the wave to the plasma. On the other hand, the term
$|\gamma_{el}|$ results in the growth of the wave, and this implies
another (stochastic) heating mechanism that also  involves  single
particle interactions with  the wave. Remark that this process is
described and \textit{even experimentally verified} \cite{14,15} for
a hot, fully ionized plasma, viz.\ in a tokamak. For drift wave
perturbations of the form $\phi(x) \cos(k_y y + k_z z - \omega t)$,
with $|k_y|\gg |k_z|$, one finds the ion particle trajectory in the
wave field from the following set of equations:
\begin{equation}
d \chi/d \tau=\Upsilon, \quad \chi\equiv k_y x,  \quad
\Upsilon\equiv k_y y, \quad \tau\equiv \Omega_i t,\label{c1}
\end{equation}
\begin{equation}
d^2 \Upsilon/d \tau^2=- \Upsilon + \left[m_i k_y^2 \phi/\left(e
B_0^2\right)\right]
  \sin\left(\Upsilon- \tau \omega/\Omega_i\right).\label{c2}
\end{equation}
It has been shown \cite{14} that stochastic heating takes place for
a large enough  wave amplitude, more precisely for $a= k_y^2
\rho_i^2 e \phi/(\kappa T_i) \geq 1$. The maximum achieved bulk ion
velocity, proportional to the wave amplitude, is given by
\be v_{max}\simeq [k_y^2 \rho_i^2 e \phi/(\kappa T_i) +
1.9]\Omega_i/k_y. \label{vm} \ee
Ideally, this all requires $ |\gamma_{el}|\geq |\gamma_{ion}|$  so
that the wave amplitude may grow and at some moment both heating
mechanisms may take place simultaneously. In the solar corona this
condition can easily and quickly be satisfied because of the almost
unlimited  range of the parallel wave number $k_z$, so that the ion
Landau damping can be made  small, i.e.\ $|\omega/k_z|\gg \vti$, and
because of the large growth rate as shown in the example in Fig.~2.

\begin{figure}
\onefigure[height=7cm,bb=45 15 305
230,clip=,width=.85\columnwidth]{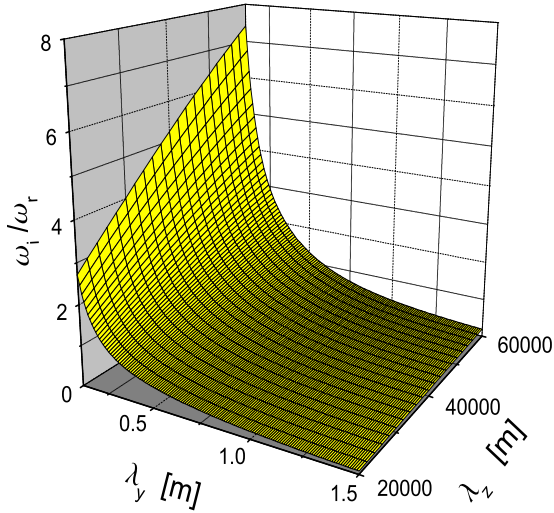} \caption{ The
growth rate (\ref{k2}) normalized to $\omega_r$ in terms of the
parallel and perpendicular
  wave-lengths. The parameters are   $B_0=10^{-2}\;$T,
   $n_0=10^{15}\;$m$^{-3}$, $L_n= [(dn_0/dx)/n_0]^{-1}=100\;$m, $T_i=10^6\;$K.} \label{fig.2}
\end{figure}

\begin{figure}
\onefigure[height=7cm,bb=45 15 290
230,clip=,width=.85\columnwidth]{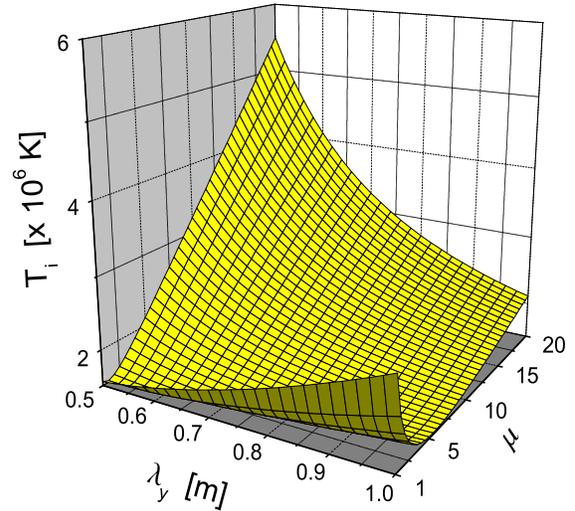} \caption{ The
temperature obtained by stochastic heating, in terms of the ion
  mass $\mu=m_i/m_p$  and the perpendicular wavelength.} \label{fig.3}
\end{figure}

In the stochastic heating process due to the drift wave,  the ions
move in the perpendicular direction to large  distances and feel the
time-varying field  of the wave due to the polarization drift $ \vec
v_p=(\partial \vec E/\partial t)/(\Omega_i B_0)$, and as a result
their motion becomes stochastic.  The polarization drift is in the
direction of the wave vector, which emphasizes the crucial
electrostatic nature of the wave in the given heating process. Also,
this   stochastic heating is  {\em highly anisotropic}, and it takes
place mainly in the direction normal to the magnetic field $B_0$
(both the $x$- and $y$-direction velocities are stochastic). The
perpendicular heating in the experiment \cite{14,15} was larger by
about a factor 3 compared to the parallel one, and it  is
exceptionally fast (see below).  At the same time, in view of the
mass difference and the physical picture given above, this heating
scenario predominantly acts on the ions.

In application to coronal magnetic structures, the indication or
`proof' that the heating really takes place would be: i)~an ion
temperature anisotropy $T_{i\bot}\gg T_{i z}$,  ii)~a possibly
higher ion temperature in comparison to electrons, and iii)~a better
heating of heavier ions.  Observations show that i)  may be taken
rather as a rule than as an exception \cite{2,3}, i.e., the
perpendicular stochastic heating is more dominant compared to the
parallel heating.  There are also numerous indications that confirm
the features ii) and iii). As an example we refer to  graphs from
Ref.~\cite{4}, where $T_e<T_H<T_{He}$ throughout  the corona and the
solar wind.

An easy way to demonstrate that the heating can take place at
various inhomogeneity scale-lengths $L_n$ (in other words in various
magnetic field or density structures in the solar corona), is to
keep the ratio $\lambda_z/L_n$ fixed. For example,  setting $L_n=
s\times 100\;$m, and $\lambda_z= s\times 40000\;$m,  where $s$ takes
values, e.g., between  1 and 1000, we calculate the frequency and
the growth rate for the fixed value $\lambda_y=0.5\;$m,  and we find
out that the ratio $\omega_r/\omega_i \simeq 1$. Note that such a
variation of $s$ may also be used to describe the natural change of
the radial density gradient with the increased altitude, in other
words the heating occurs everywhere {\em along} a given flux tube.

Assuming an initial perturbation of the order $e \phi/(\kappa
T_i)\simeq 0.01$, (i.e., $\phi=0.86\;$V)  for the parameters from
Fig.~2 (and for $\omega_i\simeq \omega_r\simeq 2.5 \cdot 10^2$ Hz,
$\lambda_y\simeq 0.5\;$m) the growth time $\tau_g=\ln 100/\omega_i$
till it becomes of the order of unity ($\phi=86\;$V) is about
$0.02\;$s. For the temperature increased by $10^6\;$K (see Table~1)
this implies a heating rate of ions of the order of $5\cdot
10^7\;$K/s, which is similar to  the heating rate obtained in the
experiments \cite{14,15}. Observe also that the magnitude of the
electric field which we are dealing with is of the same order as in
those experiments.

The stronger  heating of  heavier ions (see Fig.~3)
   can be understood from Eq.~(\ref{vm})
   and after expressing the effective temperature in terms of the ion
   mass $T_{eff}(m_i)=m_iv^2_{max}/(3\kappa)$. From the derivative
   $dT_{eff}(m_i)/dm_i>0$, it follows  that the heating increases with the
   ion mass if $k^4_y\rho^4_i ( {e\phi}/{\kappa T_i} )^2 > 1.9$.
   For $\phi=60\;$V we have the normalized temperature $T_i(\lambda_y, \mu)=0.881 + 0.057 \mu/\lambda_y^2
   + 1.78 \lambda_y^2/\mu$.

For the same  parameters as above, the maximum energy released per
unit volume is $\Sigma_{max}=n_0 m_i v_{max}^2/2=0.04\;$J/m$^3$. The
energy release rate $\Gamma_{max}=\Sigma_{max}/\tau_g \simeq
1\;$J/(m$^3$\,s) amounts to 4 orders of magnitude above the
necessary value. However, for $L_n=100\;$km (i.e., setting $s=1000$)
we obtain $\omega_i=0.13\;$Hz, $\omega_r=0.254\;$Hz, $\tau_g=
32.6\;$s, and consequently $\Gamma_{max}=1.2 \cdot
10^{-3}\;$J/(m$^3$\,s), that is about one order of magnitude above
the heating rate accepted as necessary. Similar estimates may be
done for still larger $L_n$, yet the conditions under which the
previous expressions are derived become violated and a numerical
approach is required in this case. In reality the collisions and
nonlinearity lead to the  flattening of the density profile in the
region occupied by the wave \cite{16}, that should result in the
saturation of the growth. Also, energy diffusion in the
perpendicular direction  should  act in the same way. In addition,
the drift wave is as a rule coupled to the Alfv\'{e}n wave
\cite{17}, with the coupling proportional to $k_y \rho_i$. All these
effects will more effectively act on short scales, and the actual
values for $\Gamma$ are expected to be below $\Gamma_{max}$.
Therefore, the apparently too big release of energy at short scales,
as formally obtained above, may in reality  be considerably reduced.
Clearly, more accurate estimates and more detailed description may
be obtained only numerically.

To summarize, the  proposed mechanism is based on a novel paradigm
that allows a self-consistent solution model. The heating mechanism
implies instabilities on time and spatial scales that are currently
not directly observable by space probes. However, all the effects
presented here are directly experimentally verified under laboratory
conditions. Their indirect confirmation in the context of the solar
corona seems to be also beyond doubts. This is because the {\em
consequences} of the heating process, as enlisted earlier in the
text (temperature anisotropy, better heating of heavier ions, hotter
ions than electrons), are indeed verified by satellite observations.

%%%%%%%%%%%%%%%%%%%%%%%%%%%%%%%%%%%%%%%%%%%%%%%%%%%%%%%%%%%%%%%%%%%%%%%%%%%%%%%
\begin{table}
\caption{
Plasma heating for hydrogen ions for two  perpendicular wavelengths and for two values of the
wave amplitude $\phi$.  The values in brackets are for helium.
 The maximum stochastic velocity is given by Eq.~(\ref{vm}) and the
effective temperature obtained by heating  is $ m v_{max}^2/(3
\kappa)$. We used the same starting set of parameters as earlier,
viz.\ $L_n=100\;$m, $\lambda_z=20\;$km, $n_0=10^{15}\;$m$^{-3}$,
$T=10^6\;$ K. For shorter wavelengths the heating is stronger for
helium because  the condition $k^4_y\rho^4_i ({e\phi}/{\kappa
T_i})^2> 1.9$ (see Fig.~3) is easily satisfied.  } \label{tab.1}
\begin{center}
%\begin{minipage}{50mm}
{\footnotesize\begin{tabular}{c c   c    }     % 3 columns
\hline\hline
 &$\phi= 60  $ [V] & $\phi= 80  $  [V]\\
 \hline
 $\lambda_y$ [m]  &  $ T_{eff}$ [K]  &  $ T_{eff}$ [K]  \\
\hline
     0.5  &  $1.56 \times 10^6$ ($1.91 \times 10^6$) &  $\;\;2.03  \times 10^6$ ($2.92\times 10^6$)    \\
      1  &  $2.72 \times 10^6$ ($1.56 \times 10^6$)  &  $\;\;3.06 \times 10^6$ ($2.03  \times 10^6$)      \\
\hline
     \hline
\end{tabular}}
%\end{minipage}
\end{center}
\end{table}
%%%%%%%%%%%%%%%%%%%%%%%%%%%%%%%%%%%%%%%%%%%%%%%%%%%%%%%%%%%%%%%%%%%%%%%%%%%%%%%

\acknowledgments These results  are  obtained in the framework of
the projects G.0304.07 (FWO-Vlaanderen), C~90205 (Prodex~9),  and
GOA/2009-009 (K.U.Leuven). TRACE (Fig.~1) is a mission of the
Stanford-Lockheed Institute for Space Research, and part of the NASA
Small Explorer program.

\end{document}